\documentclass{revtex4}
\usepackage{times}
\usepackage{epsfig,amssymb,amsfonts,amsmath}
\usepackage[usenames]{color}

\usepackage{amssymb}
\usepackage{amsmath}
\usepackage{amsfonts}
\usepackage{graphicx}
\usepackage{rotating}
\usepackage{ulem} 
\usepackage{cancel}
\usepackage{extarrows}

\begin{document}

\title[ ]{Reply to the comment on ``Route from discreteness to the continuum for the Tsallis $q$-entropy'' by Congjie Ou and Sumiyoshi Abe}
\author{$^{1}$Thomas Oikonomou}
\email{thomas.oikonomou@nu.edu.kz}
\author{$^{2}$G. Baris Bagci}
\email{gbb0002@hotmail.com}

\affiliation{$^{1}$Department of Physics, School of Science and Technology, Nazarbayev University, Astana 010000, Kazakhstan}
\affiliation{$^{2}$Department of Materials Science and Nanotechnology Engineering, TOBB University of Economics and Technology, 06560 Ankara, Turkey}
%
%
%

\begin{abstract}
It has been known for some time that the usual $q$-entropy $S_q^{(n)}$ cannot be shown to converge to the continuous case. In [Phys. Rev. E 97 (2018) 012104], we have shown that the discrete $q$-entropy $\widetilde{S}_q^{(n)}$ converges to the continuous case when the total number of states are properly taken into account in terms of a convergence factor. Ou and Abe [Phys. Rev. E 97, (2018) 066101] noted that this form of the discrete $q$-entropy does not conform to the Shannon-Khinchin expandability axiom. As a reply, we note that the fulfillment or not of the expandability property by the discrete $q$-entropy strongly depends on the origin of the convergence factor, presenting an example in which $\widetilde{S}_q^{(n)}$ is expandable.
%
%
\end{abstract}

\eid{ }
\date{\today }
\startpage{1}
\endpage{1}
\maketitle


The $q$-entropy in current use today reads
\begin{eqnarray}\label{Tsallis0}
S_{q}^{(n)} =\sum_{i=1}^{n}P_i\ln_{q}(1/P_i)\,,
\end{eqnarray}
where $\ln_{q} (x)$ (with $q > 0$) denotes the well-known deformed $q$-logarithm function and we always set the Boltzmann constant to unity. Note that the expression above (and also Eq. (\ref{Tsallis1}) below) yields the ordinary Shannon entropy $S^{(n)} = \sum_{i=1}^{n}P_i\ln(1/P_i)$ in the $q \rightarrow 1$ limit so that we denote Shannon entropy simply as $S$ from here on. Although still used, this discrete expression does not have a continuous counterpart as noted by Abe \cite{Abe1}. This observation has a very serious consequence, namely it limits the validity of the $q$-entropy to the discrete systems, thereby excluding any continuum use of it. However, the $q$-distribution is found in numerous continuous systems [2-9]. To overcome this difficulty, we have recently shown that the following expression can be shown to have the continuum limit \cite{our1}
\begin{eqnarray}\label{Tsallis1}
\widetilde{S}_{q}^{(n)} =n^{q-1} \sum_{i=1}^{n}P_i\ln_q(1/P_i)\,.
\end{eqnarray}

The above expression is called the discrete $q$-entropy $\widetilde{S}_{q}^{(n)}$ whereas the one in Eq. (\ref{Tsallis0}) will simply be referred to as the $q$-entropy $S_{q}^{(n)}$. Despite agreeing that the expression in Eq. (\ref{Tsallis1}) has indeed the continuum limit, Ou and Abe \cite{OuAbe} commented that it now violates the expandability property of the Shannon-Khinchin axioms \cite{Khinchine,note1} which can be stated as follows: Considering a system $A$ with the states $\Omega_A=\{A_1,\ldots,A_n\}$ and the respective probabilities $\{P_1,\ldots,P_n\}$, the expandability property says that if we add a zero-probability state, i.e., $\Omega_A'=\{A_1,\ldots,A_n,A_{n+1}\}$ with $\{P_1,\ldots,P_n,0\}$, then the entropy should not change i.e.
\begin{eqnarray}\label{exp_prop}
S^{(n+1)}(P_1,\ldots,P_n,P_{n+1}=0) = S^{(n)}(P_1,\ldots,P_n)\,.
\end{eqnarray} 
To put it simply, the expandability property demands that the zero-probability events should not change the entropy. Obviously, the expandability axiom seems very natural intuitively as it is almost the case for everything regarding the Shannon entropy.
%

Regarding now the discrete $q$-entropy in Eq. (\ref{Tsallis1}), the fulfillment of the expandability property depends on whether the convergence factor $n^{q-1}$ remains or not insensitive to the addition of an impossible event. For example, we can consider the full expression of  $\widetilde{S}_q^{(n)}$ as
\begin{eqnarray}\label{example}
\widetilde{S}_q^{(n)} = \sum_{i=1}^{n}\widetilde{P}_i [-\ln_q(P_i)]\,,
\end{eqnarray}
where $\widetilde{P}_i$ is an escort measure defined as
\begin{eqnarray}\label{escprob}
\widetilde{P}_i:= \frac{P_i^q}{\sum_{k=1}^{n}P_{k,\mathrm{equal}}^q}\,.
\end{eqnarray}
In Eqs. (\ref{example})-(\ref{escprob}) we read that the discrete $q$-entropy is the $q$-deformed uncertainty of the system averaged over the escort distribution $\widetilde{P}_i$. The distribution set $\{P_{i,\mathrm{equal}}\}_{i=1,\ldots,n}$ is nothing but the set $\{P_i\}_{i=1,\ldots,n}$ considered in the particular case of equal probabilities, i.e. $P_{i,\mathrm{equal}}=n^{-1}$, so that any change of the number of states $n$ affects in the same manner both sets and accordingly $P_{n+1}=0\;\Leftrightarrow P_{n+1,\mathrm{equal}}=0$ for an impossible $(n+1)$th event. Apparently, Eq. (\ref{example}) together with Eq. (\ref{escprob}) yields the expression in Eq. (\ref{Tsallis1}), with the latter however satisfying the expandability property, i.e.,
\begin{eqnarray}
\widetilde{S}_q^{(n+1)}(P_1,\ldots,P_n,P_{n+1}=0) = \widetilde{S}_q^{(n)}(P_1,\ldots,P_n)\,.
\end{eqnarray}

To sum up, we showed, as mentioned in the abstract, that the fulfillment of the Shannon-Khinchin axiom related to the expandability property by the discrete $q$-entropy $\widetilde{S}_q^{(n)}$ strongly depends on the origin of the convergence factor $n^{q-1}$. We provided an example where $\widetilde{S}_q^{(n)}$ satisfies the former axiom. Note that the discussion point in Ref. \cite{our1} was not to explore the origin of the aforementioned factor but to observe its existence as a necessary condition for convergence.

\begin{acknowledgments}
This research is partly supported by state-targeted program ``Center of Excellence for Fundamental and Applied Physics" (BR05236454) by the Ministry of Education and Science of the Republic of Kazakhstan and ORAU grant entitled ``Casimir light as a probe of vacuum fluctuation simplification" with PN 17098.
\end{acknowledgments}



\begin{thebibliography}{20}
\bibitem{note1} One quick way out can be writing the Boltzmann constant $k_\mathrm{B}$ as $\left( \frac{k_\mathrm{B}}{n^{q-1}}\right) $ to restore the expandability of the discrete $q$-entropy $\widetilde{S}_{q}$. However, as Ou and Abe notes, this has physically unacceptable implications. We remark that this is not the route we opted for in this reply.
\end{thebibliography}
\end{document}